\documentclass[conference]{IEEEtran}

\usepackage[normalem]{ulem}

\usepackage{epigraph}
\usepackage{url}
\usepackage{stfloats}
\usepackage{makecell}

\usepackage[strict]{changepage}

\usepackage{framed}

\usepackage{cite}
\usepackage{amsmath,amssymb,amsfonts}
\usepackage{graphicx}
\usepackage{textcomp}
\usepackage{xcolor}

\usepackage{soul}

\usepackage[noend]{algpseudocode}
\usepackage[formats]{listings}
\usepackage{caption}
\usepackage{soul}

\def\everywhere{MPI everywhere}
\def\threads{MPI+threads}

\def\multiple{\texttt{MPI\_THREAD\_MULTIPLE}}

\def\mpianytag{\texttt{MPI\_ANY\_TAG}}

\def\mpiwait{\texttt{MPI\_Wait}}

\def\mpiput{\texttt{MPI\_Put}}
\def\mpiget{\texttt{MPI\_Get}}
\def\mpiaccum{\texttt{MPI\_Accumulate}}
\def\mpiwinflush{\texttt{MPI\_Win\_flush}}

\def\noanytag{\texttt{mpi\_assert\_no\_any\_tag}}
\def\noanysource{\texttt{mpi\_assert\_no\_any\_source}}
\def\allowovertaking{\texttt{mpi\_assert\_allow\_overtaking}}

\def\naive{na\"\i ve}

\def\ie{{\it i.e.}}
\def\eg{{\it e.g.}}

\definecolor{lightGrey}{rgb}{0.93, 1, 0.93}
\newcommand{\Hilight}{\makebox[0pt][l]{\color{lightGrey}\rule[-0.45em]{\linewidth}{1.25em}}}

\newcommand{\secref}[1]{Section~\ref{#1}}
\renewcommand{\eqref}[1]{Equation~\ref{#1}}

\newcommand{\figref}[1]{Figure~\ref{#1}}
\newcommand{\coderef}[1]{Listing~\ref{#1}}

\newcommand{\tabref}[1]{Table~\ref{#1}}
\newcommand{\minititle}[1]{\textbf{{#1}}}

\definecolor{reccshade}{rgb}{0.95,0.95,1}
\definecolor{warshade}{rgb}{1,0.95,0.85}
\definecolor{greenshade}{rgb}{0.95,1,0.95}
\definecolor{darkgreen}{rgb}{0,0.5,0}
\definecolor{darkblue}{rgb}{0,0,0.7}
\definecolor{cream}{rgb}{1.0, 0.99, 0.82}
\definecolor{darkcoral}{rgb}{0.8, 0.36, 0.27}
\definecolor{darkspringgreen}{rgb}{0.09, 0.45, 0.27}

\newenvironment{concern}{%
	\MakeFramed{\advance\hsize-\width\FrameRestore}%
	\noindent\hspace{-4.55pt}
	\begin{adjustwidth}{}{7pt}%
		\vspace{2pt}\vspace{2pt}%
	}
	{%
		\vspace{2pt}\end{adjustwidth}\endMakeFramed%
}

\def\BibTeX{{\rm B\kern-.05em{\sc i\kern-.025em b}\kern-.08em
    T\kern-.1667em\lower.7ex\hbox{E}\kern-.125emX}}

\begin{document}

\title{Lessons Learned on MPI+Threads Communication}


\author{\IEEEauthorblockN{Rohit Zambre, Aparna Chandramowlishwaran}
\IEEEauthorblockA{\textit{Electrical Engineering and Computer Science} \\
\textit{University of California, Irvine}\\
Irvine, CA, USA\\
\{rzambre,amowli\}@uci.edu}
}

\maketitle

\begin{abstract}
	
	Hybrid \threads{} programming is gaining prominence, but, in
	practice, applications perform slower with it compared to the
	\everywhere{} model. The most critical challenge to the parallel
	efficiency of \threads{} applications is slow \multiple{}
	performance. MPI libraries have recently made significant strides
	on this front, but to exploit their capabilities, users must expose
	the communication parallelism in their \threads{} applications.
	Recent studies show that MPI 4.0 provides users with new
	performance-oriented options to do so, but our
	evaluation of these new mechanisms shows that
	they pose several challenges. An alternative design is MPI
	Endpoints. In this paper, we present a comparison of the
	different designs from the perspective of MPI's end-users: domain
	scientists and application developers. We evaluate the mechanisms
	on metrics beyond performance such as usability, scope, and
	portability. Based on the lessons learned, we make a case for a
	future direction.
	
\end{abstract}


\section{Introduction}
\label{sec:introduction}

The hybrid \threads{} model is gaining prominence
over the traditional \everywhere{} approach following
the evolution of modern computing architectures.
Over the last decade, the number of cores on a processor
has grown disproportionately to the growth in other on-node
resources such as memory, TLB space, and network
resources (work queues and doorbell registers)~\cite{fugaku,thakur2010mpi}.
Consequently, domain scientists have witnessed their
applications run out of memory with the memory-hungry
\everywhere{} model on large
problem sizes~\cite{pall2014tackling, bulucc2017distributed, rabenseifner2009hybrid}.
With \threads{} (\eg, MPI+OpenMP), on the other hand,
applications are able to scale to much larger problems
since the model enables users to utilize the many cores
on a processor with threads \emph{and} efficiently
share the limited on-node resources between cores
with a single process per node (or NUMA domain).
For these reasons, modern event-based frameworks (\eg{},
Legion~\cite{bauer2012legion} and YGM~\cite{priest2019you})
have not been developed with \everywhere{} from the start.

In terms of performance, however, \threads{} applications
tend to perform slower  than their \everywhere{}
counterparts in practice~\cite{jin2011high,bulucc2017distributed,higgins2015hybrid, pall2014tackling}. The reason is that
\threads{} programming raises many new
challenges, such as mitigating thread-synchronization
overheads~\cite{hetland2019paths,rodchenko2015effective, iwainsky2015many},
and preventing performance-degrading memory
accesses (\eg{}, false sharing), that are not present in
\everywhere{}. The most critical challenge, however, is the
dismal communication performance of \threads{}
applications~\cite{zambre2021logically,amer2015mpi+,thakur2010mpi,amer2019software,balaji2010fine}.
This challenge is a pressing bottleneck because most scientific
simulation campaigns run close to the strong-scaling limit where
communication has been demonstrated to occupy a significant
portion of an application's
runtime~\cite{raffenetti2017mpi,sahasrabudhe2020improving,wang2020pencil,wang2021algorithm}.

Furthermore, as an application approaches the strong-scaling
limit, the size of each individual message decreases, and the
communication performance is limited by the rate of
issuing messages rather than the network bandwidth.
In this paradigm, the interoperability of
threads with MPI is critical for the performance of an \threads{}
application.

MPI defines multiple levels of threading support
going from a highly restrictive level---only one thread will
execute---to a completely flexible level: multiple threads can execute
MPI operations in parallel (\ie, \multiple). According
to a 2017 survey of applications chosen as candidates for the
upcoming exascale systems and other application development
reports, domain scientists prefer using the flexible \multiple{}
level but do not do so currently primarily because of poor
performance~\cite{bernholdt2020survey, daily2018gossipgrad, tensorflow}.

\begin{figure*}[!t]
	\begin{center}
		\includegraphics[width=\textwidth]{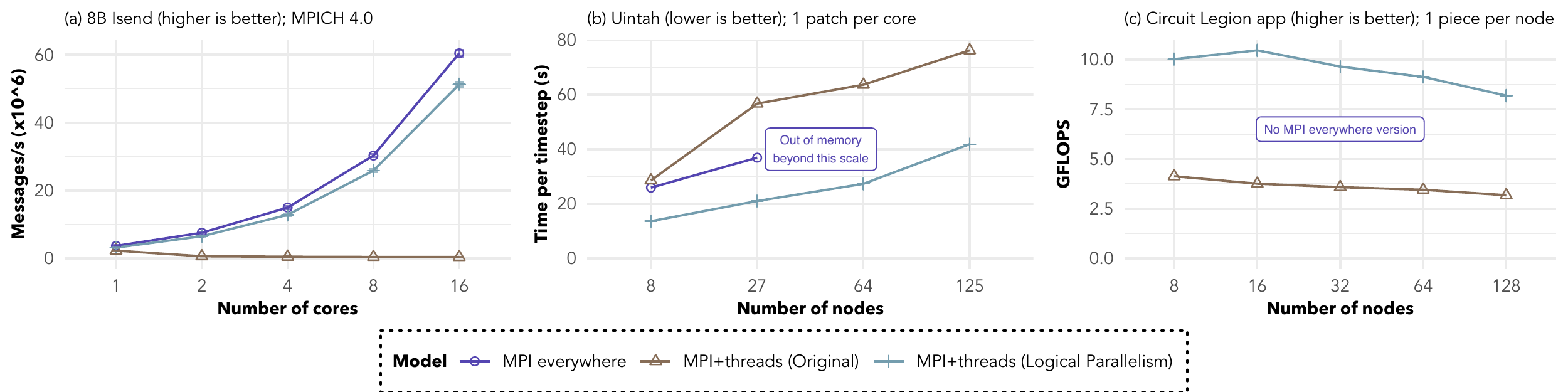}
		
		\caption{\threads{} achieves high scalability \emph{and} high performance with logically parallel communication on new MPI libraries. All \threads{} versions use \multiple{}; (a) uses Intel Skylake nodes, (b) uses Intel KNL nodes, and (c) uses Intel Broadwell nodes; the network in all experiments is Omni-Path.}
		\label{fig:sota}
	\end{center}
\vspace{-1.5em}
\end{figure*}

Recently (starting in 2019), however, MPI libraries have
made significant strides in achieving scalable multithreaded
communication performance that matches that of \everywhere{}.
\figref{fig:sota}(a) shows MPICH's support for high-speed \multiple{}
performance in its latest 4.0 release as an example.
The primary factor for this improved performance is
the MPI library's ability to map \emph{logically parallel
	communication}---operations that are not ordered
according to MPI's semantics---to the underlying network
parallelism~\cite{zambre2018scalable, zambre2020learned, patinyasakdikul2019give, intelmpi}.
With such new capabilities, applications are able to achieve
the best of both worlds---high scalability and high performance---with
\threads{} compared to \everywhere{}. Figures 1(b) and 1(c)
demonstrate the performance impact of using logically
parallel communication for traditional
stencil-style workloads (Uintah computational
framework~\cite{sahasrabudhe2020improving, zambre2021logically})
and modern data-centric workloads (Legion-based Circuit
simulation~\cite{bauer2012legion, zambre2021logically}).

In any application, the key to achieving fast \threads{} communication
is logically parallel communication. Without it, the new MPI
libraries are ineffective. So, how can domain scientists and
application developers expose
such communication parallelism?

In this regard, the MPI community first pursued
user-visible extensions to the standard in the form of
MPI
Endpoints~\cite{foster1996generalized, demaine2001generalized, dinan2013enabling}.
An endpoint represents a logically independent
stream of communication. Each endpoint is directly
addressable through an MPI rank,
making user-visible endpoints a flexible solution. 
The MPI Forum had deliberated the MPI Endpoints
proposal but ultimately suspended it on the prospect that
existing MPI objects such as communicators and
windows can expose the same level of logical
communication parallelism as user-visible endpoints.
In cases where MPI's semantics prevent users from exposing
communication independence, this school of thought
advocates the use of MPI Info hints
to relax the limiting MPI semantics which
would allow the application to expose logically parallel
communication through alternative MPI mechanisms
like tags. To analyze the performance differences
between the two approaches, recent studies map the
capabilities of both to parallel network
resources and evaluate them on applications 
from
different domains: linear system solvers, graph analytics,
astrophysics, particle physics, and  event-based runtimes.
These studies demonstrate that existing MPI
mechanisms indeed perform as well as user-visible endpoints~\cite{zambre2020learned,zambre2021logically},
and they are the basis for the introduction of new hints for
application-specific relaxation of semantics in the latest
MPI 4.0 standard.

In this paper, we show that
exposing the communication independence between
threads with existing MPI objects has various drawbacks
even with the new Info hints. Using communicators, the
most explicit existing mechanism, to expose communication
independence is quite complex even for applications with
well-structured and regular communication patterns
(\eg{}, stencil workloads). The complexity arises from
the fact that users need to ensure messages are matched
appropriately \emph{while} expressing communication parallelism.
This complexity hurts not just the productivity of the domain
scientist but also the MPI library's mapping to the underlying
network parallelism, thereby hurting performance. Relaxing
the unneeded MPI semantics through Info hints and expressing
communication parallelism through tags instead is one way
to combat this complexity, but such a solution is not portable
since the optimal mapping to the parallel network resources
is dependent on hints specific to an MPI library.

On the other hand, for the same applications, we find that
the user-visible endpoints solution is not only portable but
also straightforward because of the flexibility of its design:
each endpoint is directly addressable. Like the idea of ``ports"
first described by Foster et al.~\cite{foster1996generalized, demaine2001generalized},
endpoints do not constrain communication parallelism
information with existing MPI semantics. The concern here,
however, is the introduction of a new concept to the user.
Presumably, this concern does not hold with existing MPI
mechanisms since users are already aware of communicators,
windows, etc. However, we argue
that MPI users do not intuitively think of existing
MPI objects as means of expressing parallelism. For example,
users have historically viewed a communicator as a group
of processes. That it can double up as a means to express
logically parallel communication is a corollary of its definition.
Hence, the concern of enforcing new concepts (\ie, repurposing
existing MPI objects) on users also holds for existing MPI mechanisms.

Finally, the MPI 4.0 standard also introduced partitioned
operations as an alternative to MPI Endpoints to alleviate the
performance problem of \threads{} communication. This new
interface allows users to define a persistent message with
multiple data partitions so that each thread can drive
different partitions of a single message. Ongoing
research on this new interface demonstrates that
partitioned communication does perform better than \threads{}
communication with no logically parallel communication
(``\threads{} (Original)" in \figref{fig:sota})~\cite{dosanjh2021implementation, worley2021portable, bangalore2020portable, grant2019finepoints}, but a performance analysis against
the other two mechanisms, especially in its capabilities to utilize
the underlying network parallelism, remains to be seen.
We note, however, that the partitioned interface exhibits
certain semantic limitations which could prevent it from
matching the performance of the other approaches. For
example, threads share the single MPI request
of a partitioned message (see \secref{sec:pc}) which means
that threads would contend to access the shared request's
resources in the MPI library, or that threads would need
to synchronize to allow only one thread to poll for completion.
In either case, the application is prone to incur overheads
of contention or synchronization compared to fully
independent paths. Techniques like double or triple
buffering could partially mitigate the slowdowns caused by this
limitation, but they do not allow threads to achieve
complete independence in a manner that the other
mechanisms allow and promote.

Out of the three approaches, two have been introduced
in MPI 4.0 and at least one of them allows
users to expose complete independence between threads.
So, does this mean \emph{MPI's end users} are satisfied?
As is the case with the success of any technology, the answer
lies with the end users, the domain scientists.
We show that MPI 4.0 does
\emph{not} meet the needs of \threads{} applications, and that
it introduces
new problems. The current solutions in MPI 4.0 may be
stepping stones to alleviating the performance of \threads{}
communication, but they are not sufficient by themselves.

In this paper, we compare the strengths and
limitations of the designs (described in \secref{sec:schools})
with respect to their applicability to MPI's point-to-point,
RMA, and collective communication. Our comparison (see
\secref{sec:conversations}) is centered around the lessons
learned from evaluating a variety of
different types of \threads{} applications. Guided by both
quantitative (in \cite{zambre2020learned,zambre2021logically})
and qualitative (this paper) comparisons of the different
interfaces, in \secref{sec:winner}, we discuss
a future direction that would enable applications
to achieve not just high scalability and performance, but also
high productivity with the \threads{} programming model.

%
%
%
%
%
%
%

\section{The Three Designs for MPI+Threads Communication}
\label{sec:schools}

In this section, we describe how the three designs
can expose communication independence between
threads, and discuss their motivations and their
implementation evaluations.

\subsection{Existing MPI mechanisms}
\label{sec:existing_mechanisms}

With existing MPI objects, such as communicators,
tags, and windows, users can expose
relatively unordered operations for all types:
point-to-point, RMA, and
collectives.

\minititle{Communicators}
apply to collective and point-to-point
communication. For the former, MPI requires
collectives to be issued serially on a communicator.
Hence, users can issue collectives in parallel on 
a process only on distinct communicators.
For two-sided operations, MPI specifies a nonovertaking
order and uses the
$\langle$communicator, rank, tag$\rangle$ triplet to match
operations. Two or
more operations issued using different
communicators cannot match the same target
operation and hence imply no relative ordering;
such operations are logically parallel.
Operations that share a communicator but use
different ranks or tags, however, are not logically
parallel because of the possibility of wildcards (\eg{},
\mpianytag) on the receive side. Hence, the only
way to expose logical communication
independence for point-to-point operations
with MPI's default semantics
is through the use of multiple communicators.

\minititle{Tags with hints.} MPI 4.0 features
new Info hints that allow an
application to relax semantics it does not
need. Info hints relevant to this paper
include \allowovertaking{}, \noanytag{},
and \noanysource{}.
The first, when set, informs the MPI library that
the operations do not need to be matched in
the order that they were posted. This
Info hint is beneficial when the application
requires wildcards but does not require
the MPI library to maintain the order of
matching. With the nonovertaking order relaxed,
two or more send operations
using different
tags are logically parallel even if they use
the same communicator and address the same
target process.
Because of wildcards, however, similar receive
operations are not logically parallel. If the
application does not require wildcards, the domain
scientist can set the other two hints to relax
the wildcard's constraints. Without any
wildcards, two or more operations (both send
and receive) that use the same communicator,
address the same process, but use different
tags can never match with the same target
operation. Hence, such operations are logically
parallel.

\minititle{Windows} apply
to MPI's RMA operations.
By default, MPI maintains program
order only for its atomic operations (\eg{},
\mpiaccum) originating from the same source
and targeting the same memory location on the same
window. Otherwise, both atomic and nonatomic
RMA operations (\eg{}, \mpiput) on different
windows are unordered. Although
nonatomic operations are logically parallel
in any case, users need to be wary of mixing synchronization
and initiation operations in parallel on the same
window. For example, if one thread is waiting inside
\mpiwinflush{} and another continuously issues
\mpiget{} operations, the first thread might block
indefinitely. To overcome such cases and
explicitly expose parallelism for any type of RMA
operation, users have the option of using distinct
windows for different threads.

By mapping the communication independence
exposed by the above mechanisms to the underlying
network parallelism, recent research demonstrates that
\threads{} applications can indeed achieve scaling
communication throughput that matches that of \everywhere{}~\cite{zambre2020learned,zambre2021logically, patinyasakdikul2019give}

\subsection{User-visible Endpoints}
\label{sec:endpoints}

Researchers had initially proposed to extend the MPI
standard to introduce user-visible MPI endpoints~\cite{dinan2013enabling}.
With a new API, users can create communicators with
multiple endpoints (see \figref{code:ep_api}).
This API creates a new communicator (context ID)
from an existing one, \texttt{parent\_comm},
and provides \texttt{my\_num\_ep} number of
handles to the new communicator. Each handle is
addressable with a distinct rank. For all operations, users
would specify the local endpoint to use using one of the
returned handles (\texttt{new\_comm\_handles}) and address a target endpoint using
the endpoint's rank, a global index of the endpoint,
making endpoints a flexible interface.
One could then use an endpoints communicator to create
endpoints for other MPI objects such as windows and
files~\cite{mpiendpointsproposal}.

\begin{figure}[!b]
	\vspace{-1.75em}
	\begin{lstlisting}
MPI_Comm_create_endpoints(parent_comm, my_num_ep, info, new_comm_handles)
	\end{lstlisting}
	\caption{
		API to create a communicator with multiple endpoints.
	}
	\label{code:ep_api}
\end{figure}

Each endpoint takes on the semantics of an MPI
rank. Like messages originating from different processes,
messages from different endpoints are unordered and hence 
logically parallel. If the user maps each thread to a
distinct endpoint, then all threads are directly
addressable. Given their flexible interface, user-visible
endpoints represent the upper bound in expressing the
communication parallelism available in an application.
Several efforts show scaling multithreaded communication
throughput with user-visible endpoints~\cite{sridharan2014enabling,dinan2014enabling,holmesintroducing}.

One of the notions on which the endpoints proposal was
suspended was that some networking hardware may
not be able to optimize the creation of new network
addresses (for new endpoints) after initializing the MPI
library~\cite{dinan2014enabling}.
We note, however, that the new MPI libraries
have addressed this problem by creating a pool of network
resources during the initialization phase~\cite{patinyasakdikul2019give,zambre2020learned}.
Such implementations then map logical entities like
endpoints to physical network resources.

\subsection{Partitioned Communication}
\label{sec:pc}

MPI 4.0's partitioned communication interface allows
users to specify a persistent message with multiple partitions.
Each partition contributes to a single message, and the
contributions could occur in parallel from multiple
threads. Users define the operation's characteristics 
(\eg, number of partitions, tag, etc.)
outside the critical path, and then contribute the individual
partitions of the message whenever a thread is ready
to communicate (see \figref{code:pc_apis}).
MPI 4.0 contains APIs to describe standard-mode send
and receive operations only, but the idea can extend to other
modes of point-to-point operations and even RMA and collective operations~\cite{holmes2021partitioned}. These
extensions will require their own set of APIs to define
the partitioned communication equivalent for each operation.

\begin{figure}[htbp]
\vspace{-0.75em}
	\begin{lstlisting}
MPI_Psend_init(buf, num_partitions, count, datatype,
               dest, tag, comm, info, request)
MPI_Precv_init(buf, num_partitions, count, datatype,
               source, tag, comm, info, request)
MPI_Pready(partition, request)
MPI_Parrived(partition, request, flag)
	\end{lstlisting}
	\caption{New APIs to create and use standard mode send and receive partitioned
		communication operations.
	}
	\label{code:pc_apis}
\end{figure}

Partitioned communication was introduced to combat
the message-matching overheads in multithreaded
communication. Message matching is a costly serial
operation~\cite{schonbein2020low}. If $n$ threads
use the same communicator (``\threads{} (Original)"
in \figref{fig:sota}), the overhead of message
matching grows by $O(n)$. Since partitioned operations
share a persistent message, they incur a
message matching overhead of only $O(1)$ for $n$ threads
driving the multiple partitions of the message.
Research implementations have demonstrated
performance benefits of partitioned operations
especially for large partitions even with older MPI
libraries that do not capitalize on parallel network
resources~\cite{grant2019finepoints, dosanjh2021implementation}.
Partitioned operations are suited to benefit from
the capabilities of the new MPI libraries (multiple
partitions could map to distinct network resources),
but such a study has not been conducted yet. More
important, how partitioned operations compare
to the other mechanisms of exposing logically
parallel communication where message matching
is not a concern (due to a distinct matching engine per
communication channel~\cite{patinyasakdikul2019give,zambre2020learned})
remains to be seen.

\section{Application-Centric Comparisons of\\The Three Designs}
\label{sec:conversations}

In this section, we discuss how the different
designs (see \secref{sec:schools})
compare against each other with respect to two key metrics:
ease of use (which reflects the productivity of domain
scientists), and applicability to different MPI operations (which
measures the scope of the designs). We map
the communication patterns of key applications
to the different design choices to make such 
a comparison. In the process of doing so,
we collaborate with application developers from a
variety of institutions including
University of Utah (stencil communication in the
hypre linear solver used by Uintah~\cite{sahasrabudhe2020improving}), 
Maison de la Simulation (stencil communication
in Smilei~\cite{derouillat2018smilei}), Pacific Northwest
National Laboratory (graph communication in Vite~\cite{ghosh2018scalable}),
University of California, Irvine (stencil communication
in Pencil~\cite{wang2020pencil}), HPE (RMA
communication in WOMBAT~\cite{mendygral2017wombat}),
and Argonne National Laboratory (Legion's MPI
backend~\cite{legionmpibackend}). We organize
our discussion below by the lessons learned from
comparing the different designs.

\subsection{Point-to-point communication}
\label{sec:pt2pt}

\minititle{Mechanism 1: Communicators.}
Using distinct communicators is the most explicit
way to express logically parallel communication
for point-to-point operations with MPI's default
semantics (see \secref{sec:existing_mechanisms}).
We discover, however, that communicators
pose several challenges: complexity, high resource usage,
and lack for flexibility for irregular and dynamic
communication patterns. We detail our lessons
learned below.



\begin{concern}
	\emph{Lesson 1:} Exposing logically parallel
	communication with communicators is a complex task
	due to its matching requirements.
\end{concern}

\begin{figure}[!t]
	\centering
	\includegraphics[width=0.3\textwidth]{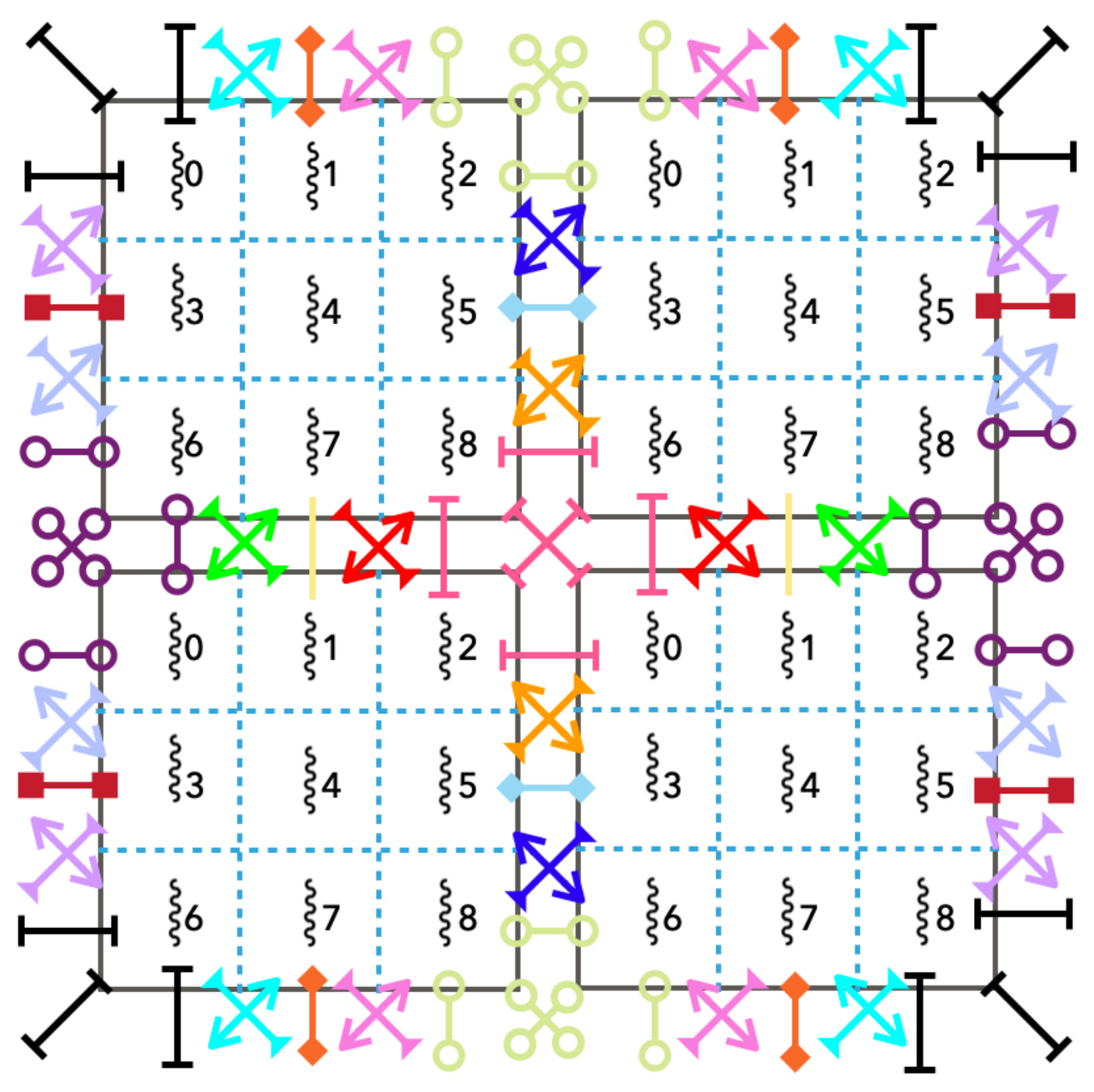}
	\caption{Ideal communicator usage for a 2D 9-point stencil
		(stencils are the core kernels in hypre~\cite{sahasrabudhe2020improving},
        Smilei~\cite{derouillat2018smilei}, and Pencil~\cite{wang2020pencil}).
		Each box represents a process with 9 threads. Each thread
		has 1 patch. Each color-shape combination represents a
		communicator. Numbers represent thread IDs.}
	\label{fig:2d_stencil_comm}
	\vspace{-1em}
\end{figure}

To understand the first lesson, let us consider a relatively
simple example of a static (communication pattern of each thread is
fixed) 2D 9-point stencil. \figref{fig:2d_stencil_comm}
shows the ideal communicator usage---minimum number of
communicators with all of the available parallelism exposed---for
such a communication pattern. For a given direction
of communication, we have as many communicators as there 
are communicating threads on the edge (a plane in 3D) since
the operations of the threads are independent. The threads
on a corner, however, use a single communicator for all directions
since their operations for the different directions occur serially.
The mapping of communicators to threads is not
the same on each process. For example, thread 7 of the bottom-left
process in \figref{fig:2d_stencil_comm} must use a communicator
for its north-south communication that is different from the
communicator that thread 7 on the top-left rank uses for the
same north-south direction. This difference in communicators
prevent threads 1 and 7 on a process from using the
same communicator and serializing their communication. In
other words, given a map of communicators for the threads of
a given process, the map for other processes can be derived
by mirroring the map along the change in cartesian coordinates
of the process.
\coderef{code:stencil_using_comms} shows a 2D
MPI+OpenMP 5-point stencil that exposes communication
parallelism using communicators. For simplicity,
it does not optimize
communicator usage for corner threads. Lines 23--26 
demonstrate the mirroring of communicator assignment
to threads. This mirroring idea extends to the
diagonal exchange in 9-point stencil as well:
the user would need to extend lines 12--17 to create 4
additional sets of communicators---2 diagonals along the
NS boundaries and 2 along the EW boundaries---each
containing as many threads as there on the edge.
Combining the mirroring strategy with the
optimization to use a single communicator for corner
threads will achieve the map illustrated in
\figref{fig:2d_stencil_comm} but at the cost of 
further complexity. The user would need to extend
lines 23--24 to mirror the assignment of communicators
to threads along both axes for all directions of exchange. 
Diagonal communicator sets require another
dimension of mirroring that is exemplified by thread
1 on the bottom left process (see \figref{fig:2d_stencil_comm}) using the same
communicator for its NE exchange as does thread 1
on the bottom right process for its NW exchange.
Optimizing for reduced communicator usage is important
for efficient use of network resources (see Lesson 3 below).

\begin{concern}
	\emph{Lesson 2:} Using communicators to
	expose communication parallelism is not intuitive.
\end{concern}


Continuing on the 2D 9-point stencil
example,
the intuitive approach to expose
communication parallelism is
to create as many communicators as there are threads and
then use communicator $i$ for thread $i$'s send operations and
communicator $j$ for thread $i$'s receive operations where $j$ is
the thread id of the remote thread that thread $i$ is receiving from.
This usage of communicators is correct, but it exposes \emph{only half}
of the available parallelism. The communication of adjacent threads
on an edge occur in parallel
but the operations of threads on opposite edges use the
same communicator. For example, in \figref{fig:2d_stencil_comm},
thread 1's send operation uses communicator 1, which 
thread 7 also uses for its receive operations. 

\begin{concern}
	\emph{Lesson 3:} Communicators have high network
	resource requirements to expose communication parallelism.
\end{concern}

\begin{figure}[!t]
	\begin{minipage}{0.49\textwidth}
		\begin{lstlisting}{mpi3}
recv_from(proc_rank, tag, comm, *req, tid) :
  if (need_mpi_op(tid)) :
$\Hilight$    MPI_Irecv(proc_rank, tag, comm, req)
  else : /* use shared memory */
send_to(proc_rank, tag, comm, *req, tid) :
  if (need_mpi_op(tid)) :
$\Hilight$    MPI_Isend(proc_rank, tag, comm, req)
  else : /* use shared memory */
void main() :
  /* px*py process grid with tx*ty local thread grid */
  //  Create communicators to expose parallelism
$\Hilight$  for (i = 0; i < tx; i++) :
$\Hilight$    MPI_Comm_dup(COMM_WORLD, &ns_comm_a[i])
$\Hilight$    MPI_Comm_dup(COMM_WORLD, &ns_comm_b[i])
$\Hilight$  for (i = 0; i < ty; i++) :
$\Hilight$    MPI_Comm_dup(COMM_WORLD, &ew_comm_a[i])
$\Hilight$    MPI_Comm_dup(COMM_WORLD, &ew_comm_b[i])
  /*Neighbor process ranks: n_rank,s_rank,e_rank,w_rank*/
#pragma omp parallel num_threads(N_THREADS)
  {
    /* Coords in local tx*ty thread grid: tid_x, tid_y */
    // Choose the right communicator to use
$\Hilight$    n_comm = (ry%2) ? ns_comm_b[tid_x] : ns_comm_a[tid_x]
$\Hilight$    s_comm = (ry%2) ? ns_comm_a[tid_x] : ns_comm_b[tid_x]
$\Hilight$    e_comm = (rx%2) ? ew_comm_b[tid_y] : ew_comm_a[tid_y]
$\Hilight$    w_comm = (rx%2) ? ew_comm_a[tid_y] : ew_comm_b[tid_y]
    for (iter = 0; iter < niters; iter++) :
      recv_from(n_rank, tag_ns, n_comm, &reqs[0], tid)
      /* recv_from [s_|e_|w_]rank with [s_|e_|w_]comm */
      send_to(n_rank, tag_ns, n_comm, &reqs[4], tid)
      /* send_to   [s_|e_|w_]rank with [s_|e_|w_]comm */
      MPI_Waitall(8, reqs)
      /* Compute after halo exchange */
  }
		\end{lstlisting}
\end{minipage}
\captionof{lstlisting}{
	2D 5-point stencil using MPI communicators.
}
\label{code:stencil_using_comms}
\vspace{-1.25em}
\end{figure}

Even if the domain scientist achieves the ideal communicator
usage, the number of communicators required to express
communication parallelism is much higher than the minimum
number of parallel channels required by the communication
pattern itself. Such high resource requirements are concerning
on today's many-core architectures. Consider the communication pattern
of real-world stencil applications which is typically a 3D 27-point
stencil (\eg{}, hypre). Such applications decompose their
domain into cubical patches. If $[x,y,z]$ represents
the cubic arrangement of threads in an MPI
process, the least number of communicators needed to express
all of the available logical communication parallelism is
\(
2xy + 2yz + 2xz +
8(xy+ yz + xz - 1) +
4(xz + yz - z) + 4(xy + yz - y)  + 4(xy + xz - x)
\).
The first three terms represent the directions
perpendicular to the 6 faces, the fourth term represents
the 8 corner diagonals, and the last three terms
represent the edge diagonals. In terms of parallelism
alone, however, the minimum number of parallel
communication channels required is 
$xyz - (x-2)(y-2)(z-2)$ which is the number
of threads communicating inter-node. If we consider
a 64-core processor (\eg{}, AMD EPYC Rome), the minimum
number of communicators required to express
communication parallelism is 808 (1 process per node
with $[4,4,4]$ threads per process) which is over
$14\times$ higher than the minimum number of
communication channels required by the 3D 27-pt
stencil communication pattern. For this same
pattern, other mechanisms use only as many
resources as the communication pattern requires
(see ``Mechanism 3" below).
The resource inefficiency of communicators can even
hurt performance on some networks where the number
of network hardware resources is limited (\eg{}, 160
hardware contexts on Omni-Path~\cite{hfi_guide}).
Prior studies show hypre's communication time is
over $2\times$ higher with communicators than with
other mechanisms on Omni-Path~\cite{zambre2021logically}.
In such a scenario, the domain scientist is expressing
all of the available communication parallelism
and the application is using all of the network resources,
but the observed performance benefit may not be
as expected because of contention on the limited number
of network resources which includes the software
overheads of thread synchronization to access shared
network queues~\cite{zambre2018scalable}.

\begin{concern}
	\emph{Lesson 4:} Overloaded definitions of
	communicators can lead to mismatch in expected
	mapping to the underlying network parallelism.
\end{concern}

A communicator has historically been viewed
as a group of processes or as a
means to isolate matching of messages. That
it can double up as a means to express parallelism
is a corollary of its definition. The
multiple functions of a communicator can lead to a
mismatch in expected mapping to the underlying network
parallelism. For example, an application can initially create a set
of communicators for grouping different
processes and later use communicators to express
parallelism. The MPI library underneath cannot differentiate
between the two and could end up allocating a significant
portion of the underlying
network resources to the communicators used for grouping
different sets of processes, leaving fewer network resources
to map to for logical-parallelism-oriented communicators.
MPI libraries can prevent this type of mismatch in
expected mapping by introducing  hints that allow an
application to inform the library when it is creating
communicators for the purposes of expressing logically
parallel communication. But such hints would be
implementation-specific.

\begin{concern}
	\emph{Lesson 5:} The matching semantics of communicators 
	limit communication parallelism for irregular
	and dynamic communication patterns.
\end{concern}

\begin{figure}[!t]
	\centering
	\includegraphics[width=0.3\textwidth]{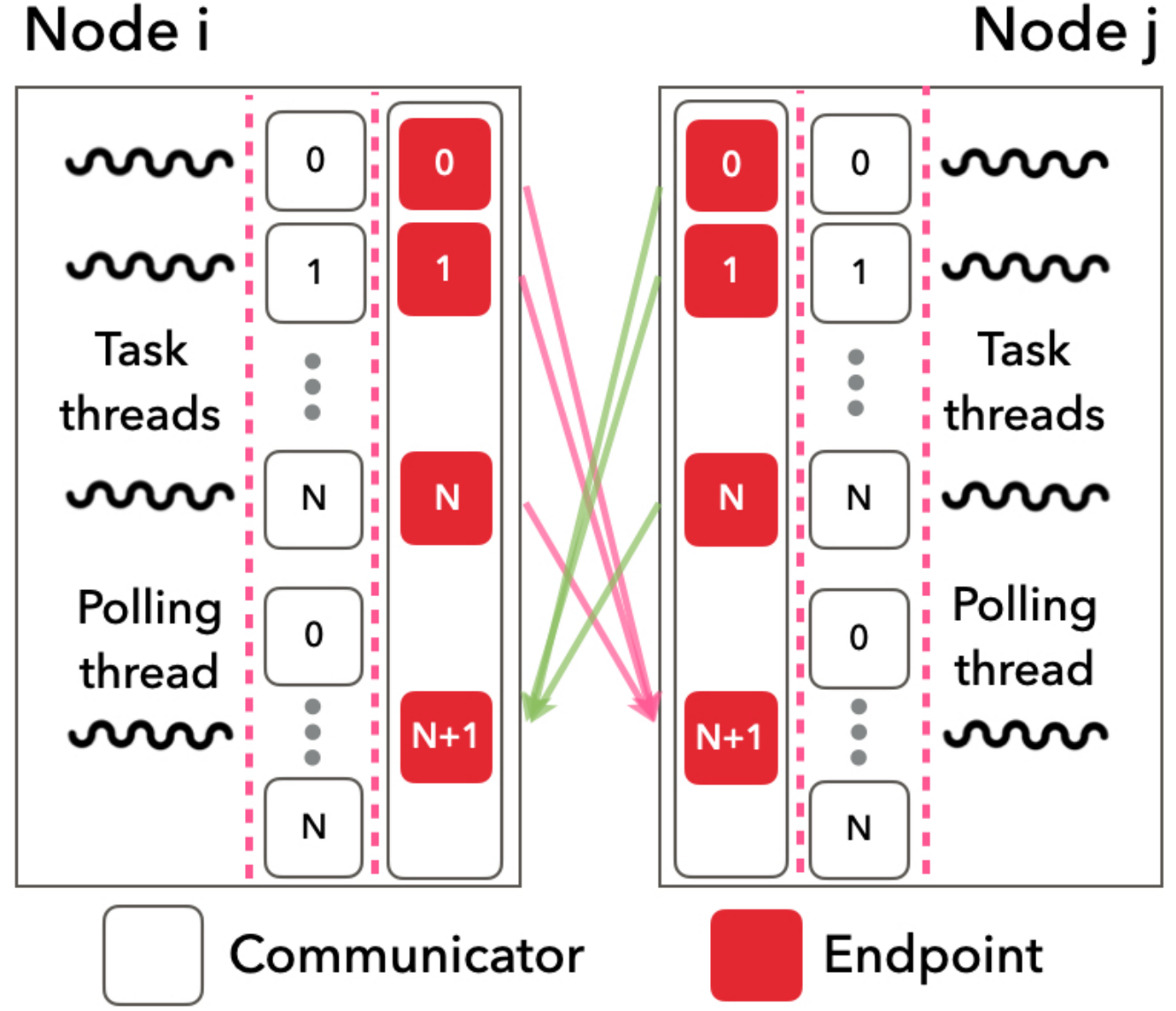}
	\caption{Communicator vs. endpoints for Legion's runtime that
		implements an event-based system using
		multithreaded point-to-point communication.}
\label{fig:legion_comm}
\vspace{-1em}
\end{figure}

As shown in \figref{fig:2d_stencil_comm}
and lines 23--26 in \coderef{code:stencil_using_comms},
applications must ensure that the sending and receiving
threads use the same communicator. This matching
constraint is limiting for applications where
the communication neighborhood of a thread changes
over time, as it does in graph (\eg{},
Vite~\cite{gawande2022towards}) and adaptive mesh
refinement applications. This constraint also
holds for applications running
	on modern task-based frameworks
that exhibit irregular communication patterns.
\figref{fig:legion_comm} portrays this limitation
for Legion~\cite{bauer2012legion} applications. Legion's
event-based runtime maintains a receiving polling thread
per node to process incoming requests from the task threads
on other nodes. The multiple task threads on a node can issue operations
using distinct communicators, but the polling thread is forced
to iterate over the communicators to process all incoming messages.
Thus, on a single node, the polling thread conflicts with the
communicators of the task threads. The polling thread relies
on wildcards, and hence using partitioned operations for this
communication pattern is not straightforward. With endpoints,
on the other hand, the polling thread can use a distinct endpoint,
use wildcards, and satisfy matching requirements.
Prior evaluations show that Legion's polling thread processes
events $1.63\times$ slower with communicators than with 
endpoints~\cite{zambre2021logically}.

\minititle{Mechanism 2: Tags with hints.}
The new Info hints in MPI 4.0 allow domain scientists
to use tags instead of communicators
to express logical parallelism in applications that
do not use certain MPI semantics (see \secref{sec:existing_mechanisms}).
Tags bypass some of the challenges with communicators,
but introduce new ones. We describe the lessons learned
from using tags below.

\begin{concern}
	\emph{Lesson 6:} Using tags for communication
	parallelism is intuitive.
\end{concern}

Most \threads{} applications that use
\multiple{} already encode thread IDs into the tags of
their communication to differentiate operations that
target different threads on the same target process
(\eg{}, hypre and Smilei), indicating that
domain scientists intuitively think of tags as a means
of expressing logical parallelism. Hence, the approach of using
tags requires the least amount of
changes to existing applications. These changes would
only be in the form of creating a new communicator
with Info hints that relax unneeded MPI semantics
(see \coderef{code:using_tags}).

\begin{concern}
	\emph{Lesson 7:} Achieving optimal multithreaded
	communication performance with tags is tedious.
\end{concern}

Even
though tags and communicators have the same
matching constraints, tags can provide more
information. Consider the MPICH library that
features multiple virtual communication interfaces
(VCIs: network communication channels mapping
to distinct network hardware contexts)~\cite{zambre2020learned},
and the hypre library that encodes the IDs of the
sending and receiving threads into the tag along
with other application-related information.
The thread IDs in the tag provide information
about which local and remote VCI to use.
But if MPICH does not know which bits of the tag
encode communication parallelism information,
then hypre is at the mercy of how MPICH hashes
the tags into the multiple VCIs allocated to the
communicator. Achieving the optimal mapping to
VCIs requires hypre to inform MPICH which bits
encode the sender's thread ID, which bits encode
the receiver's thread ID, and  how to map the bits
to the underlying VCIs. For example, with a one-to-one
mapping, MPICH can use the
sender-thread-bits to map to a VCI on the host
process and the receiver-thread-bits to decide
which VCI to target on the remote process.
Such intricate use of tags requires domain scientists to learn about
the Info hints that are specific to an MPI library
(see lines 6--9 in \coderef{code:using_tags}).

\begin{figure}[!t]
	\begin{minipage}{0.49\textwidth}
		\begin{lstlisting}{Pt2pt}
/* Existing THREAD_MULTIPLE stencil apps */
#pragma omp parallel num_threads(N_THREADS)
{
  app_tag = src_tid << (NUM_TID_BITS + NUM_APP_BITS)
          | dst_tid <<  NUM_APP_BITS
          | app_tag; 
  MPI_Send(dest_proc, tag, app_comm);
}
		\end{lstlisting}
	\end{minipage}\hfill
	\begin{minipage}{0.49\textwidth}
		\begin{lstlisting}{RMA}
/* Leveraging parallelism info in tags of existing apps */
//(1) Relax unneeded MPI semantics with MPI 4.0 hints
MPI_Info_set(info,"mpi_assert_no_any_tag",   "true");
MPI_Info_set(info,"mpi_assert_no_any_source","true");
//(2) Achieve optimal mapping with MPI library hints
MPI_Info_set(info,"mpich_num_vcis",            N_THREADS);
MPI_Info_set(info,"mpich_num_tag_bits_vci", NUM_TID_BITS);
MPI_Info_set(info,"mpich_place_tag_bits_local_vci","MSB");
MPI_Info_set(info,"mpich_tag_vci_hash_type","one-to-one");
MPI_Comm_dup_with_info(app_comm, info, &tag_par_app_comm);
#pragma omp parallel num_threads(N_THREADS)
{
  ... // (Tag encoding same as above)
  MPI_Send(dest_proc, app_tag, tag_par_app_comm);
}
		\end{lstlisting}
	\end{minipage}
	\captionof{lstlisting}{
		Exposing logically parallel
		communication through existing MPI tags in applications
		 like Smilei and hypre.
	 }
	\label{code:using_tags}
	\vspace{-1em}
\end{figure}

\begin{concern}
	\emph{Lesson 8:} Programs that use tags for
	communication parallelism may not be portable
	across MPI libraries.
\end{concern}

As discussed above, achieving a high level
of control and flexibility with tags requires a careful
use of implementation-specific Info hints.
It is likely that different MPI implementations
will support tag-based and communicator-based
mechanisms of exposing parallelism in different
ways (VCIs are specific to MPICH; CRIs are specific to Open MPI~\cite{patinyasakdikul2019give}). Since HPC
application developers are geared towards
performance-oriented codes, they are compelled
to adopt Info hints to optimally expose logically
parallel communication to an MPI implementation.
Hence, existing MPI mechanisms can result in
reduced portability of codes which is highly undesirable.

\begin{concern}
	\emph{Lesson 9:} Encoding communication
	parallelism in tags is limited by their existing
	use cases.
\end{concern}

End
users already use MPI tags for
application-related information. Since the number
of bits in a tag is limited, an application may not be
able to encode further parallelism information into
the tag. Encoding parallelism information with
lesser-than-ideal number of bits is bound to hurt
performance.
Although we have not encountered first-hand an
application that faces such a problem, others
have reported running into tag-overflow issues
on prominent applications (\eg{}, SNAP~\cite{snaptag},
Smilei~\cite{smileitag}, and MITgcm~\cite{gcmtag}).
Such reports indicate that applications already use
a large portion of the tag space; encoding
parallelism information into tags exacerbates the
tag-overflow problem.

\minititle{Mechanism 3: User-Visible Endpoints.}
Given their flexible interface (see \secref{sec:endpoints}),
endpoints combat the various concerns 
associated with communicators and tags to expose
logically parallel point-to-point operations.
Below, we delineate the lessons learned from mapping
endpoints to different communication patterns.

\begin{concern}
	\emph{Lesson 10:} Endpoints are intuitive to use.
\end{concern}
Endpoints are an easier alternative to express
communication parallelism even for patterns such as
a 3D 27-point stencil
since each local endpoint can flexibly
address any other endpoint through a global endpoint rank.
They are more intuitive
to use than communicators because application
developers are innately familiar with the semantics
of traditional MPI ranks. Users 
express communication parallelism by
communicating between endpoints as they
do for MPI ranks in \everywhere{} programming.
Lines 17--20 in \coderef{code:stencil_using_eps}
shows this MPI-everywhere like addressing in a
2D MPI+OpenMP 5-pt stencil that exposes
communication parallelism with endpoints.
In
fact, endpoints provide a level of flexibility beyond \everywhere{}:
threads are not bound to an endpoint. In other words,
endpoints do not enforce an association between threads and
the data they work on; a thread is free to
use any endpoint at any time. Thus, endpoints map
well to tasking frameworks like OpenMP Tasks~\cite{de2018ongoing}.

\begin{concern}
	\emph{Lesson 11:} Endpoints distinguish between
	matching and parallelism information and thus
	apply seamlessly to all types of communication
	patterns.
\end{concern}

Unlike existing MPI mechanisms, user-visible
endpoints separate the task of expressing
communication parallelism from the task of
matching operations. Thus, using endpoints is
straightforward even for irregular communication
patterns, such as those of Legion applications (see \figref{fig:legion_comm}).
Endpoints can flexibly adapt to dynamically
changing communication neighborhoods because
threads can address new remote endpoints
while using the same local endpoint
(lines 22--25 in \coderef{code:stencil_using_eps}). 
Additionally, endpoints do not suffer from
overloaded definitions of existing MPI objects. Users do not
need to compromise on the application's existing use
of communicators or tags.

\begin{figure}[!t]
	\begin{minipage}{0.49\textwidth}
		\begin{lstlisting}{endpoints}
recv_from(ep_rank, tag, comm, *req, tid) :
  if (need_mpi_op(tid)) :
$\Hilight$    MPI_Irecv(ep_rank, tag, comm, req)
  else : /* use shared memory */
send_to(ep_rank, tag, comm, *req, tid) :
  if (need_mpi_op(tid)) :
$\Hilight$    MPI_Isend(ep_rank, tag, comm, req)
  else : /* use shared memory */
void main() :
  /* px*py process grid with tx*ty local thread grid */
  // Create as many endpoints as there are threads
$\Hilight$  MPI_Comm_create_endpoints(COMM_WORLD, N_THREADS, info,
$\Hilight$                            &comm_ep)
#pragma omp parallel num_threads(N_THREADS)
  {
    /* Coords in local tx*ty thread grid: tid_x, tid_y */
$\Hilight$    n_ep = n_rank*N_THREADS + tx*(ty-1) + tid_x;
$\Hilight$    s_ep = s_rank*N_THREADS + tid_x;
$\Hilight$    e_ep = e_rank*N_THREADS + tid_y*tx  + tx-1;
$\Hilight$    w_ep = w_rank*N_THREADS + tid_y*tx;
    for (iter = 0; iter < niters; iter++) :
      recv_from(n_ep, tag_ns, comm_ep[tid], &reqs[0], tid)
      /* recv_from [s_|e_|w_]ep with comm_ep[tid] */
      send_to  (n_ep, tag_ns, comm_ep[tid], &reqs[4], tid)
      /* send_to   [s_|e_|w_]ep with comm_ep[tid] */
      MPI_Waitall(8, reqs);
    /* Compute after halo exchange */
  }
		\end{lstlisting}
	\end{minipage}
\captionof{lstlisting}{
		2D 5-point stencil using MPI endpoints.
}
\label{code:stencil_using_eps}
\vspace{-1.25em}
\end{figure}

\begin{concern}
	\emph{Lesson 12:} Endpoints lead to efficient
	resource usage and provide optimal
	mapping information without sacrificing
	portability.
\end{concern}

By creating an endpoints communicator,
users explicitly inform the MPI library
that the new communicator is
for the purposes of exposing communication parallelism.
This information is baked into the API
unlike the implementation-specific solution of
using hints with communicators or tags.
This is why endpoints do not suffer from
high resource requirements either. In the prior example
of executing hypre's 3D 27-point stencil on a 64-core
processor, users need to create only as many endpoints
as there are communicating threads, which is 56, 
14.4$\times$ fewer than that required by
communicators. Furthermore, the endpoints
mechanism directly provides the MPI
library with all the information needed to
optimally map to the underlying network
resources unlike the tag-based mechanism
which requires the application to inform the MPI
library about the specific tag bits that encode
logical parallelism information. Since the
optimal mapping information can be derived
from a standardized interface, applications
would be portable across MPI implementations.



\minititle{Mechanism 4: Partitioned Communication.}
MPI 4.0 introduced the new semantics of partitioned
operations for multithreaded point-to-point communication (see \secref{sec:pc}).
Like endpoints, partitioned operations promote portability,
but their semantics can limit exposure of parallelism in
irregular and dynamic communication patterns. We
expand on our lessons learned with partitioned operations
below.

\begin{concern}
	\emph{Lesson 13:} Partitioned
	operations do not overload existing definitions,
	and they promote portability of codes.
\end{concern}

Partitioned operations provide the same benefits
as user-visible endpoints on two fronts: (a) they do
not overload the definitions of existing MPI objects
and hence minimize the mismatch in expected mapping to
network parallelism; (b) they promote
portability of applications across MPI implementations
by their standardization. Unlike endpoints that
build on top of existing MPI semantics, partitioned
operations introduce new semantics. Given that research
on partitioned operations is ongoing, the usability of 
partitions remains to be seen.

\begin{figure}[!t]
	\begin{minipage}{0.49\textwidth}
	\begin{lstlisting}{partitioned}
test_recv_from(part_id, req, *rx_flag, tid) :
  if (need_mpi_op(tid)) :
$\Hilight$    MPI_Parrived(part_id, req, rx_flag)
  else : /* use shared memory */
    *rx_flag = 1
send_to(part_id, req, tid) :
  if (need_mpi_op(tid)) :
$\Hilight$    MPI_Pready(part_id, req)
  else : /* use shared memory */
void main() :
  /* px*py process grid with tx*ty local thread grid
     Each thread assigned to a tile_x*tile_y tile
     Neighbor process ranks: n_rank,s_rank,e_rank,w_rank*/
  // Create partitioned operations for parallel exchanges
$\Hilight$  MPI_Precv_init(n_rx_buf, tx, tile_x, MPI_DOUBLE, n_rank,
$\Hilight$                 tag_ns, COMM_WORLD, info, &reqs[0])
$\Hilight$  /* MPI_Precv_init for s_rank with tx partitions
$\Hilight$     MPI_Precv_init for [e_|w_]rank with ty partitions */
$\Hilight$  MPI_Psend_init(n_tx_buf, tx, tile_x, MPI_DOUBLE, n_rank,
$\Hilight$                 tag_ns, COMM_WORLD, info, &reqs[4])
$\Hilight$  /* MPI_Psend_init for s_rank with tx partitions
$\Hilight$     MPI_Psend_init for [e_|w_]rank with ty partitions */
$\Hilight$  MPI_Startall(8, reqs)  // Activate all operations
#pragma omp parallel num_threads(N_THREADS)
  {
    /* Coords in local tx*ty thread grid: tid_x, tid_y */
    for (iter = 0; iter < niters; iter++) :
      send_to(tid_x, reqs[4], tid)
      /* send_to s_rank      with partition tid_x
         send_to [e_|w_]rank with partition tid_y */
    while (!n_rx_flag ||!s_rx_flag || !e_rx_flag
           || !w_rx_flag) :
       test_recv_from(tid_x, reqs[0], &n_rx_flag, tid)
       /* test_recv_from s_rank      with part tid_x 
          test_recv_from [e_|w_]rank with part tid_y */
    /* Compute after halo exchange */
$\Hilight$#pragma omp single
$\Hilight$    { // Complete the requests before issuing next parts
$\Hilight$      MPI_Waitall(8, reqs)
$\Hilight$    } // required implicit thread barrier
  }
	\end{lstlisting}
\end{minipage}
	\vspace{-0.25em}
	\captionof{lstlisting}{
			2D 5-point stencil using MPI partitioned operations.
	}
	\label{code:stencil_using_partitions}
	\vspace{-1.1em}
\end{figure}

\begin{concern}
	\emph{Lesson 14:} Partitioned semantics prevent
	threads from being completely independent.
\end{concern}

	The fundamental limitation of partitioned communication
	is that all threads (driving the multiple
	partitions in parallel) share the same MPI request.
	So, all threads would either contend on the MPI library's
	resources of the shared request or coordinate with each other
	to allow only a
	single thread to poll for the completion of a partitioned
	operation.
    In either case, threads will incur contention
	or synchronization overheads before issuing their
	partition of the next message.
Lines 37--40 in \coderef{code:stencil_using_partitions}
shows this synchronization requirement for a 2D
MPI+OpenMP 5-pt stencil that exposes communication
parallelism using partitioned operations.
	Application developers
	could use multiple partitioned operations (\eg{}, double
	buffering) to dampen the overhead resulting from the
	semantic limitation, but they cannot eliminate
	them in a manner like the other two designs can.
    The implicit point of contention in the partitioned interface
	makes an application prone to the known high overheads
	of thread
	synchronization~\cite{baker2012scaling,hetland2019paths,rodchenko2015effective, iwainsky2015many}. It is not yet clear how the synchronization
    limitations of partitioned operations can be mitigated in modern
    applications where threads operate independently of each other~\cite{sahasrabudhe2021optimizing,treichler2014realm}.

\begin{concern}
	\emph{Lesson 15:} Persistence of partitioned
	operations prevent them from being used
	in dynamic and irregular communication patterns.
\end{concern}	
	
In dynamic communication where the destination
of a message is not known apriori, using partitioned
operations is a challenge since they are persistent by
definition (lines 15--23 in \coderef{code:stencil_using_partitions}).
Also, partitioned receive operations
cannot use wildcards. Modern task-based runtimes
(\eg{}, Legion~\cite{treichler2014realm} and YGM~\cite{priest2019you}),
however, have irregular communication patterns and
rely on wildcards in their polling threads.
Mapping partitions to the
communication pattern in
\figref{fig:legion_comm} is challenging.

%

\subsection{RMA communication}
\label{sec:rma}

One-sided RMA have no matching semantics. Here, existing
MPI mechanisms and user-visible endpoints are
equally straightforward to use, but they each have
unique concerns. The efficacy of partitioned
operations for one-sided communication is yet
to be studied.

\begin{concern}
	\emph{Lesson 16:} Where the semantics of existing MPI
	mechanisms limit the exposure of logically parallel atomic
	operations, those of endpoints achieve optimal mapping
	of operations to the underlying network parallelism.
\end{concern}

\begin{figure}[!t]
	\begin{center}
		\includegraphics[width=0.49\textwidth]{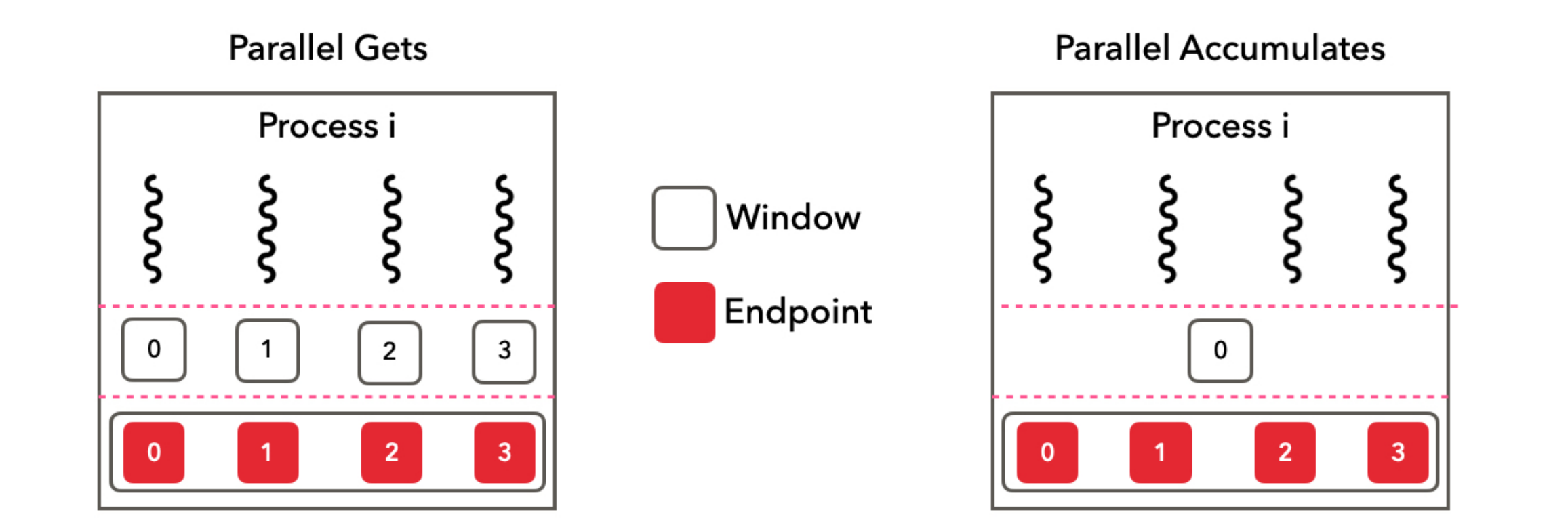}
	\end{center}
	\caption{\threads{} communication parallelism in NWChem's
		block-sparse matrix multiplication.
	}
	\label{fig:bspmm_comm}
	\vspace{-1em}
\end{figure}

Using windows to
expose communication parallelism constrains the
parallelism information with MPI's atomicity
semantics. This constraint limits the user from 
explicitly exposing logically parallel atomic
operations within a single window even when the
application does not need them to be ordered.
Consider NWChem's get-compute-update pattern
for its block-sparse matrix multiplication~\cite{si2015scaling,zambre2020learned}
where a thread uses \mpiget{} operations
to retrieve the tiles it needs and, after the multiplication,
uses an \mpiaccum{} operation to update the
destination tile (see \figref{fig:bspmm_comm}).
The \mpiaccum{} operations in a multithreaded
process must use a single window
for correct atomicity. Even though these
parallel operations are
independent, users have no
way to explicitly expose this parallelism. The best they
can do is relax MPI's
ordering constraint (\texttt{accumulate\_ordering=none})
and rely on the MPI library's hashing policies
to map operations to parallel network channels.
Any hashing policy, however, is prone
to collisions and will prevent some operations
from mapping to distinct network channels.
With user-visible endpoints, on the other
hand, users can use multiple endpoints
within a single window to expose communication
parallelism \emph{and}
maintain atomicity.


\begin{concern}
	\emph{Lesson 17:}
	There exists preconceived notions in the MPI community
	about endpoints being direct handles to network
	resources.
\end{concern}

A common misconception in the MPI community
is to view endpoints as
direct handles to network resources. This
concern holds not just for RMA operations but
also for point-to-point operations.
As a result, the endpoints
design is sometimes incorrectly regarded as
a way for MPI libraries to dump the
responsibility of managing network resources
on the domain scientist which would in turn
reduce the portability of applications.
One explanation for this concern is the usage of the
term ``endpoints," which is typically associated with
``network endpoints." The fact that user-visible
endpoints were introduced for the purposes of utilizing
network parallelism is likely to have exacerbated
the misconception. User-visible endpoints are \emph{not}
handles to network resources, rather they are a means 
to flexibly express communication parallelism. Their
usage is separate from the MPI library's task to map
the exposed parallelism to the underlying parallel
network resources. With the endpoints solutions, 
applications would create as many endpoints as
there are streams of logically parallel communication.
The MPI library would then funnel the streams of
logically parallel communication on distinct
network hardware contexts depending on their
availability.

\subsection{Collective communication}
\label{sec:coll}

Research towards implementing
collectives in a hybrid \threads{} environment
has primarily revolved around hierarchical
algorithms where threads first implement the
collective (\eg{}, allreduce) amongst themselves
and then one thread on each node participates
in the internode collective~\cite{li2013numa,zhou2020collectives}.
Recent studies using a fast \threads{} library (Intel
MPI 2019~\cite{intelmpi}), however, demonstrate performance
benefits when multiple threads drive a collective
in parallel~\cite{wende2019openmp,boyle2017accelerating}.
Although existing MPI semantics require collectives
on a communicator to be issued serially,
applications may partition the
collective-data of a process across threads and issue
parallel collectives on the different data segments
using a distinct communicator for each
thread (\eg{}, VASP collectives observe a speedup
of over 2$\times$ with such an approach~\cite{wende2019openmp}).

\begin{concern}
	\emph{Lesson 18:} Users need to perform the intranode
	portion of a collective with existing MPI mechanisms,
	but not with endpoints or partitioned operations.
\end{concern}

\begin{figure*}[!t]
	\begin{center}
		\includegraphics[width=0.75\textwidth]{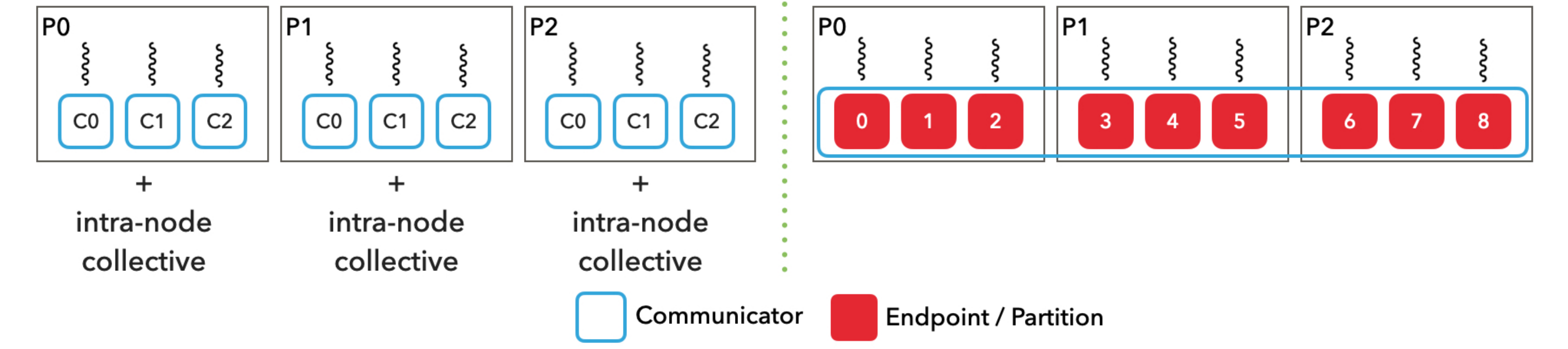}
	\end{center}
	\caption{Existing mechanisms vs. user-visible endpoints for collectives
		(\eg{}, multithreaded Allreduce in VASP~\cite{wende2019openmp}).
	}
	\label{fig:collective_compare}
	\vspace{-1em}
\end{figure*}

With existing MPI mechanisms, users need to perform
the intranode portion of the collective (see
\figref{fig:collective_compare}). For example, in an
allreduce collective, the user needs to perform
a reduction step after all threads have completed the
internode part of the allreduce. With user-visible endpoints
or partitioned operations,
on the other hand, the collective is only one step---all
threads participate
in a collective of the same communicator through
different endpoints or partitions. The MPI library then
conducts both the internode and intranode parts of
the collective before returning from the operation.
Although the performances of the
two approaches are yet to be compared, we note that, from a
design perspective, the endpoints and partitioned
approaches are better because they do not force the
user to manually handle the intranode portion of
a collective. Arguably, shared-memory
programming models feature direct support for
collectives between threads; for example,
OpenMP supports a reduction operation through
compiler directives. But such support does not
apply to all types of
collective communication that MPI features
(\eg{}, \texttt{MPI\_Alltoall}).
A \naive{} implementation for such cases is likely
to hurt performance for high thread counts (relevant
on existing and upcoming many-core processors).
Efficiently implementing a collective
is not a trivial task; researchers
have spent numerous efforts into optimizing
collectives~\cite{hoefler2014energy,chan2007collective}.
Manually implementing
tree-style and bucket algorithms is a tedious task that
could instead be handled by the MPI library as is the case
in the user-visible endpoints and partitioned operations
designs.

\begin{concern}
	\emph{Lesson 19:} Unlike existing MPI mechanisms
	and partitioned operations, user-visible endpoints
	lead to duplicated buffers on a node for some
	collectives.
\end{concern}

The interface of user-visible endpoints
results in duplication of data per node in cases
where the result of the collective is the
same across all ranks participating in the operation
(\eg{}, allreduce, broadcast, etc.). The destination
buffers of the endpoints on a single process 
contain the same values (like \everywhere{} where
each process features a copy of the resulting buffer)
when only one such buffer
is needed since all threads can directly read from the
single buffer. With communicators, on the other hand, such
duplication does not exist. For example, the user can perform
an intra-node reduction
into a single buffer that all threads can read from.
The partitioned communication
interface overcomes the duplication issue with
endpoints.
Each partition of a partitioned collective would
be different sections of the input buffer. With
threads driving distinct partitions,
the MPI library would implement both the intranode
and internode portion of a collective, and each process
would host only one buffer that contains the result
of the collective. We note, however, that we have yet
to identify a case where an \everywhere{} application
has run out of memory solely because of duplication
of a collective's resulting data. Hence, the duplication
of data in collectives with user-visible endpoints is not
as concerning as the duplication of domain-level
data that has caused applications to run out of memory
with \everywhere{}. Furthermore, the duplication
does not hold for collectives where the result of the
collective is different for different ranks (\eg{}, alltoall).

%

\subsection{Heterogenous computing environments}
\label{sec:heterogenous_computing}

The discussion in this paper so far has largely been
in the context of (CPU-initiated) \threads{} communication
because the performance-oriented studies of the designs
in the context of accelerators are yet to be conducted.
Nevertheless, we briefly discuss how the different designs
compare and apply to heterogeneous computing
environments. Today's distributed applications that use
accelerated computing have to largely rely on the
control transferring back to the CPU from the accelerator
(\eg{}, GPU) before exchanging data with remote nodes.
The system and runtime overheads (\eg{}, GPU kernel
launch latencies), however, limit the parallel efficiency
of an accelerated application. One way to combat the
scalability issue is to initiate communication
from the accelerator. Technologies like NVSHMEM~\cite{hsu2020initial}
and ROC\_SHMEM~\cite{hamidouche2020gpu} support
GPU-initiated communication for the OpenSHMEM~\cite{chapman2010introducing}
programming model. GPU-initiated (point-to-point)
MPI communication, however, remains an open problem.
Executing MPI's matching engine
on the GPU is known to be expensive~\cite{klenk2017relaxations}.

\begin{concern}
	\emph{Lesson 20:} Partitioned operations
	provide lightweight interfaces for device-initiated
	communication; the other two designs do not.
\end{concern}

Out of the three designs, partitioned operations are best
suited for high-speed device-initiated point-to-point operations.
Through its non-critical-path \texttt{P[send|recv]\_init}
APIs, partitioned communication enables most of the serial
overhead of setting up a low-level network message to be
executed on a low-latency CPU core (before kernel launch)
rather than a high-latency GPU compute unit. GPU thread
blocks
would then trigger or check for arrival of partitions with
the lightweight \texttt{Pready} and \texttt{Parrived}
operations~\cite{pcaccelbindings}. Nonetheless, the
limitations of partitioned operations described in Lessons
14 and 15 apply to heterogeneous computing scenarios as
well---program control would need to return back to the CPU
(\eg{}, to execute an \mpiwait{}) before the GPU
can issue the next partitions of a message. Such repeated
transfers of control will re-introduce device runtime
overheads that device-initiated communication aims to
address in the first place. Extensions that enable MPI
operations to be enqueued into accelerator's work queues
(similar to Nvidia's NCCL runtime~\cite{nccl}) may reduce such
runtime synchronization overheads. These extensions,
however, could apply to existing MPI objects and
user-visible endpoints too.

Another way for applications to combat the device
runtime overheads that hurt scalability is to use
persistent GPU kernels that offload
communication operations to the faster CPU cores
through lightweight atomics or flags. How such an
approach compares to device-initiated communication
remains to be seen. Such application-level techniques
are promising given the move towards system
architectures with tightly integrated CPUs and
GPUs where the latency to communicate between
the two types of PUs will diminish~\cite{gracehopper, vijayaraghavan2017design}. 

Furthermore, the benefits of device-initiated
communication either compared to or in conjuction
with techniques that leverage smart NICs or network
hardware tag matching remains to be seen. All in
all, the lessons from this paper remain relevant
for heterogeneous computing environments
moving forward.

\section{Meeting the Needs of Domain Scientists}
\label{sec:winner}

\emph{"Rule of thumb for UX: More options, more problems."}

\hfill--- \emph{Scott Belsky}

\begin{table*}[!b]
	\centering
	\caption{Summary of design choices to expose logically parallel communication
		(TBD: to be defined).}
	\label{tab:design_summary}
	\begin{tabular}{  | r | c | c | c |}
		\hline
		\textbf{Operation} & \textbf{Existing MPI mechanisms} & \textbf{User-Visible Endpoints} & \textbf{Partitioned Communication} \\ \hline
		Point-to-point & Communicators or tags & Endpoints & Partitioned point-to-point APIs \\ \hline
		RMA & Window(s) & Endpoints & Partitioned RMA APIs (TBD) \\ \hline
		Collective & \makecell{Communicators + \\ user-driven intra-node collective} & Endpoints & Partitioned collective APIs (TBD) \\ \hline
	\end{tabular}
\end{table*}

\tabref{tab:design_summary} shows a summary
of the design choices to expose logically parallel
communication for different types of MPI
operations. With
existing MPI mechanisms, users have to be aware
of a multitude of options since each mechanism
does not uniformly apply to all communication
types and patterns. Domain scientists need to be aware
about the mechanisms that become available when
hints relax different semantics.
Furthermore, each
mechanism poses unique challenges: using
communicators is hard because of MPI's
matching semantics; the optimal use of tags is
highly dependent on MPI implementation-specific
hints; windows may not allow users to
optimally expose the available communication
independence.

The new partitioned communication interface
also poses challenges to the user. The
new semantics of partitions
and the large expansion in
MPI's API space indicate a multitude of
options for the user to learn about and choose
from. The interface is challenging to use 
for dynamic and irregular communication 
patterns especially those that use wildcards.
More important, the semantic limitations of the
interface prevent the user from achieving complete
independence between threads.

One way to combat the limitations of the interfaces
in MPI 4.0
is to design
an abstraction on top of MPI that allows users
to seamlessly expose communication independence
in a user-friendly manner. The abstract layer would
then use MPI 4.0 mechanisms underneath with
MPI-implementation specific hints where needed (analogous
to how different interconnects support
the Open Fabrics Interfaces
(OFI)~\cite{grun2015brief} API through their own
OFI providers). The challenge of such an
abstraction is the design of an interface that
applies to all communication patterns.
The interface of user-visible endpoints is
in fact an example of such a general abstraction; other
forms remain to be investigated.
But, more important, the implementation of 
any abstraction faces
the semantic constraints
of both existing MPI mechanisms (\eg, matching semantics
of communicators) and partitioned operations
(\eg, no wildcards).

In contrast, with user-visible endpoints,
the interface that has not been standardized yet,
users need to be aware of only one mechanism: \emph{endpoints},
which applies uniformly to all types of MPI
operations.
Endpoints provide a flexible,
straightforward interface for users to
express logically parallel communication in a way
that they are already familiar with (\ie{}, using
ranks). The concern with user-visible endpoints
is a misimpression among domain experts 
about what endpoints
represent. A reason for this is that the terminology
in the MPI Endpoints proposal is oriented
towards the community of MPI library developers.
Since the ultimate goal of the proposal is to
aid the domain scientist to express logically
parallel communication, it is imperative that
the proposal be user-facing.

To resonate with domain experts, we
suggest rebranding\footnote{Rebranding techniques
	have proven to be successful with many technologies
	(\eg, Android, Airbnb, etc.).} the proposal to \emph{MPI
	Rankpoints} since it emphasizes that
users can create multiple MPI ranks within a
process. The goal is to educate
and reinforce the understanding
that \emph{rankpoints} are not handles to
network resources, rather they are a flexible,
straightforward means of expressing parallelism
that promotes portability of applications. While it
requires one new API---\texttt{MPI\_Comm\_create\_rankpoints}---it
prevents the limitations of existing MPI mechanisms
and partitioned operations.
Rebranding is more
than just a change in the name of the proposal.
It requires a concerted effort by the MPI
community in re-education through venues
such as presentations and tutorials at flagship
conferences and workshops.
\section{Concluding Remarks}
\label{sec:conclusion}

\emph{"People ignore design that ignores people."}

\hfill--- \emph{Frank Chimero}

\threads{} is a critical model to program the many-core
processors of the current HPC clusters and the upcoming
exascale systems. It is imperative that applications perform
productively with it. The key to achieving high performance
with \threads{} requires effort from both MPI library
developers and application developers. Recently, MPI
libraries have made significant strides in this regard, and
now the ball is in the court of domain scientists to expose
communication parallelism to utilize the new fast \threads{}
libraries. Domain scientists, however, face many programming
challenges (Lessons 1--5, 7--9, 14--16, 18) with the
designs present in MPI 4.0 with respect to exposing the
communication independence between threads. These
solutions have their own merits (Lessons 6, 13, 20), but
we show that they do not meet the needs of key communication
patterns sufficiently.
The MPI Rankpoints alternative,
on the other hand, elegantly addresses the various limitations
of the designs in MPI 4.0. MPI Rankpoints has its own challenges
(Lessons 17 and 19), but its benefits (Lessons 10--12, 16, 18)
prove to be a seamless option for domain scientists given
that the design applies generally to all communication patterns.
The lessons in this paper show that MPI Rankpoints, or
a solution that is as flexible, warrants continued
consideration as a viable solution for the \threads{}
programming model.

\section*{Acknowledgments}
We thank the following individuals for
their discussions and viewpoints on the
different mechanisms of exposing
logically parallel MPI communication:
Damodar Sahasrabudhe from the University
of Utah, Hengjie Wang from the University
of California, Irvine, Julien Derouillat from
Maison de la Simulation, Peter Mendygral
from HPE Cray, Sayan Ghosh and Mahantesh
Halappanavar from Pacific Northwest National
Laboratory, Roger Pearce from Lawrence
Livermore National Laboratory, and Hui Zhou
and Pavan Balaji from Argonne National Laboratory.
This work is supported by the National Science
Foundation (NSF) under the award number
1750549.

\pagebreak

\bibliographystyle{abbrv}
\small
\bibliography{bib/refsused}

\end{document}